\documentclass{elsart4-1}

\usepackage{graphicx}

\usepackage{amssymb}

\usepackage[english,francais]{babel}

\newtheorem{e-proposition}[theorem]{Proposition}

\newtheorem{e-definition}[theorem]{Definition\rm}

\renewcommand{\theequation}{\arabic{equation}}
\def\og{\leavevmode\raise.3ex\hbox{$\scriptscriptstyle\langle\!\langle$~}}
\def\fg{\leavevmode\raise.3ex\hbox{~$\!\scriptscriptstyle\,\rangle\!\rangle$}}

\begin{document}

\centerline{Physics or Astrophysics/Header}
\begin{frontmatter}


\selectlanguage{english}
\title{History and results of the Riga dynamo experiments}


\selectlanguage{english}
\author[riga]{Agris Gailitis},
\ead{gailitis@sal.lv}
\author[fzd]{Gunter Gerbeth},
\ead{G.Gerbeth@fzd.de}
\author[fzd]{Thomas Gundrum},
\ead{Th.Gundrum@fzd.de}
\author[riga]{Olgerts Lielausis},
\ead{mbroka@sal.lv}
\author[riga]{Ernests Platacis},
\ead{platacis@sal.lv}
\author[fzd]{Frank Stefani\corauthref{cor1}}
\corauth[cor1]{Corresponding author.}
\ead{F.Stefani@fzd.de}

\address[riga]{Institute of Physics, University of Latvia, LV-2169 Salaspils 1,
Latvia}
\address[fzd]{Forschungszentrum Dresden-Rossendorf, P.O. Box 510119, D-01314 Dresden,
Germany}


\medskip
\begin{center}
{\small Received *****; accepted after revision +++++}
\end{center}

\begin{abstract}
On 11 November 1999, a self-exciting magnetic eigenfield was detected for the
first time in the Riga liquid sodium  dynamo experiment. We report on 
the long history leading
to this event, and on the subsequent experimental campaigns which 
provided a wealth of data on the kinematic and the saturated regime
of this dynamo. The present state of the theoretical understanding
of both regimes is delineated, and some comparisons with other laboratory dynamo
experiments are made.

\vskip 0.5\baselineskip

\selectlanguage{francais}
\noindent{\bf R\'esum\'e}
\vskip 0.5\baselineskip
\noindent
{\bf Historique et r\'{e}sultats des exp\'{e}riences dynamo de Riga.}                             
Le 11 novembre 1999, un mode propre magn\'{e}tique auto-excit\'{e} \'{e}tait
d\'{e}tect\'{e} pour la premi\`{e}re fois dans l'exp\'{e}rience dynamo au sodium liquide de Riga.    
Nous relatons le long
historique qui a conduit \`{a} cet \'{e}v\'{e}nement et pr\'{e}sentons les campagnes 
exp\'{e}rimentales qui ont suivi et qui ont produit de tr\`{e}s nombreuses
donn\'{e}es sur les r\'{e}gimes cin\'{e}matique et satur\'{e}
de cette dynamo. L'\'{e}tat actuel de compr\'{e}hension th\'{e}orique des deux r\'{e}gimes
est esquiss\'{e}e 
et des comparaisons avec d'autres exp\'{e}riences dynamo sont fournies.                              
{\it Pour citer cet article~: A. Gailitis, G. Gerbeth, Th. Gundrum, O. Lielausis, E. Platacis, F. Stefani,
 C. R. Physique  (2008).}

 \vskip 0.5\baselineskip                    
\noindent{\small{\it Mots-cl\'es~:} Dynamo; Champ magn\'{e}tique; Sodium liquide}}
\end{abstract}
\end{frontmatter}


\selectlanguage{english}
\section{Introduction}
\label{}
It is widely believed that almost any flow of a conducting liquid will give rise to 
self-excitation of a magnetic field provided that, first, 
the flow topology is not too simple (e.g. a flow in one direction or a purely rotational flow)     
and, second, the so-called magnetic
Reynolds number is large enough. This dimensionless 
number $Rm=\mu_0\sigma L V$, which is defined as the 
product
of the magnetic permeability $\mu_0$ and the electrical conductivity $\sigma$
of the fluid and 
the typical length 
scale $L$ and the typical  velocity scale $V$ of its flow, measures the
ratio of diffusive time scale
to kinematic time scale in the induction equation for
the magnetic field $\bf B$:
\begin{eqnarray}
\frac{\partial {\bf B}}{\partial t}=\nabla \times ({\bf v} \times {\bf B})+
\frac{1}{\mu_0 \sigma} \nabla^2 {\bf B}
\end{eqnarray}

While this number is large in many astrophysically relevant flows, simply 
due to their huge spatial extension,  it requires
significant effort to produce a flow with $Rm\sim 10...100$ in the 
laboratory.
The simple reason for this is that the electrical conductivity
of liquid metals, even in the optimum case of liquid sodium, does not
exceed a value of 10$^7$ S/m, which
leads to a product $\mu_0 \sigma\sim 10$ s/m$^2$. Hence, in order to get
an $Rm\sim 10$, it requires a product of fluid dimension and flow velocity
of $V L \sim 1$ m$^2$/s.
It is this large number, in connection with the precautions that
are indispensable for the safe handling of sodium, which had prevented
dynamo experiments for a long time.

There are many possibilities to organize a dynamo-active flow in a
vessel, and quite a number of them  have  been 
tried in experiment. By now, experimental dynamo science is 
completely international with strong activities in Latvia, Germany, 
France, US, and Russia. 
These attempts, only three of which were successful in showing dynamo action,      
have been summarized in a number of recent review papers
\cite{CHOSSAT,MAHYDRAEDLER,RMP,TILGNER,MOMOMO,PETRELIS},                                       
and some of them are also described in the present special issue.
Therefore, in this paper we will restrict our attention to the Riga dynamo 
experiment, 
with only a few comparative side-views on other experiments.

\section{The history of the Riga dynamo experiment}

The idea of the Riga dynamo experiment originates from one of the 
simplest dynamo concepts that was investigated by Ponomarenko in 
1973 \cite{PONO}.
The Ponomeranko dynamo consists of one conducting rigid rod which undergoes a
spiral motion within - and in gliding contact with - a medium of the 
same conductivity
that extends infinitely in radial and axial direction.
People working on planetary, stellar, or galactic dynamos might be sceptical
about the physical relevance of such a system. The first answer 
to them is that this dynamo represents a ''elementary cell'' of a number
of more complicated dynamos, and in this function it deserves particular
attention  in its own right. The second answer
is that possibly some natural systems  work in a similar manner
as the Ponomarenko dynamo. One promising candidate is  the ''double helix nebula''
which was detected recently in the outflow from the galactic centre
\cite{MORRIS}. The basic idea that cosmic jets could work as a Ponomarenko-like dynamo
was already discussed in an early paper by Shukurov and Sokoloff \cite{SHUKUROV}.

Soon after its invention by Ponomarenko, the dynamo was further analyzed by
Gailitis and Freibergs \cite{GAFR76} who found a remarkably low                %
critical magnetic Reynolds number for the convective instability of 17.7 
(with the radius taken as the defining length scale).                                           
An essential step towards an experimental realization
was the consideration of  a straight back-flow concentric to the inner helical flow,
which converts  the convective instability into an absolute one \cite{GAFR80}.

Based on this early analytical and numerical work, a first experimental attempt  
was undertaken in 1986 by Gailitis et al. \cite{GAI87}.
In contrast to the later
Riga dynamo experiment, the central helical flow in this experiment
was actualized by a spiral flow guide (the ''helical labyrinth'') 
at the entrance of flow.
The main experimental result  was the observation of
a significant amplification of an externally applied magnetic field.
Unfortunately, the experiment had to be  stopped due to some construction problems 
leading to strong mechanical vibrations, and so it is not known
if this dynamo would have been able to show self-excitation.                                 

With this experience in mind, it was decided to change slightly the concept of 
the experiment by replacing the ''helical labyrinth'' 
by a propeller and some guiding blades,  
restricted to the inlet of the central tube. Later we will see that the lack of 
mechanical constraints along the 3 m distance from the
propeller region to the bottom qualifies this dynamo as a markedly {\it fluid}   
dynamo, in which the saturation mechanism of the magnetic field strongly 
relies on the deformability of 
the velocity field.         
In addition to the concentric back-flow channel a 
third cylinder with sodium at rest
was attached in order to further reduce the critical $Rm$ 
due to improved boundary conditions for the electric currents.

A good deal of work was devoted to the optimization of the geometry and 
the details of the flow structure. The main dimensions, i.e. length and 
the radii of the
three cylinders were optimized in \cite{GAI96}. A further idea of
optimization relied on the expectation that a flow with maximum helicity (at fixed 
kinetic energy) should represent a sort of optimum for dynamo action to occur.
The calculus of variation for the corresponding problem in cylindrical geometry
leads to Bessel functions of the zeros and first order for the axial and 
the azimuthal
velocity, respectively. It turned out, that radial flow profiles of this kind
lead indeed to a minimum for the critical magnetic Reynolds number \cite{KLUWER}. 
For people familiar with Taylor relaxation 
and the reversed field pinch \cite{TAYLORRMP}, in which
a turbulent plasma produces a  magnetic field of the same Bessel function
shape, this might appear as an intriguing duality of velocity optimization and
magnetic-field self-organization.

Besides the general problems in setting up a large scale liquid sodium facility, the
realization of the desired flow structure was a time-consuming
process in particular. Most of the work was done by G. Will, M. Christen and H. H\"anel
at the Technical University of Dresden, were a 1:2 water dummy model                          
of the Riga sodium facility was built and utilized for flow measurements and optimization 
\cite{WILL}. While the 
propeller 
geometry was
fixed at an early stage,  many numerical and experimental effort was
needed to find a configuration of pre- and post-propeller vanes suitable to 
shape the flow  as close as possible to the desired Bessel function profiles.

Between 1995 and 1999, the dynamo facility was installed 
at the Institute of Physics in Salaspils, Latvia (actually, the name                            
''Riga dynamo experiment'', which has been chosen for  a better recognizability, 
is not correct, 
since Salaspils is an independent town
25 km eastward of Riga). The
main components of the facility 
and the structure of the simulated (and experimentally widely confirmed)
magnetic eigenfield are shown in Fig. 1.

\begin{figure}[tbp]
\centering
\includegraphics[width=15cm]{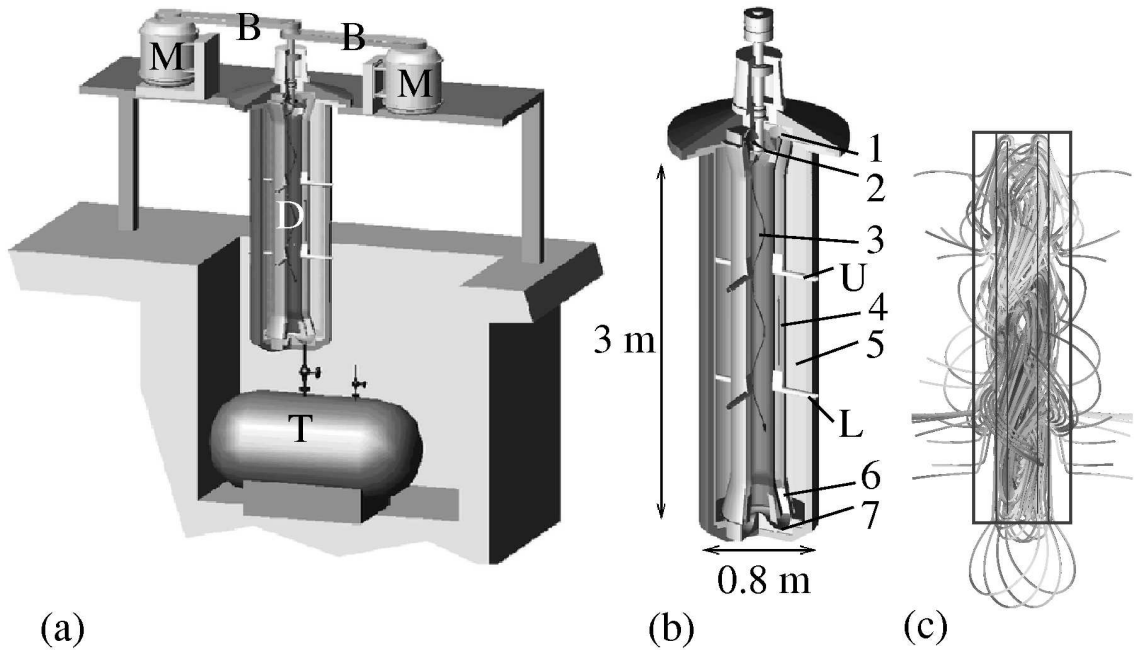} %
\caption{The Riga dynamo experiment and its eigenfield.
(a) Sketch of the Riga dynamo facility.
M - Two 100 kW motors. B - Belts from the motors to the propeller shaft. D - Central dynamo            
module. T - Sodium tank.
(b) Details of the central dynamo module. 1 - Upper bending region; 2 - Propeller;
3 - Central helical flow region; 4 - Return-flow region; 5 - Outer sodium
region; 6 - Guiding vanes for straightening the flow in the return flow. 
7 - Lower bending region. At approximately 1/3 (L) and 2/3 (U) of the dynamo height there are
four ports for various magnetic field, pressure and velocity probes.
(c) Simulated structure of the magnetic eigenfield                                                              
in the kinematic regime.}
\end{figure}

After all preparations had been done, a first experimental campaign took place
on 10-11 November 1999. Before an experiment starts, sodium is usually
heated up to  300$^{\circ}$C and pumped slowly  through the central module 
for about 24 hours
in order to achieve good electrical contact between sodium and the internal 
stainless steel walls.
It was planned to decrease the temperature to approximately 150$^{\circ}$C since the
electrical conductivity and hence the magnetic Reynolds number
increase with decreasing temperature. During this cooling-down process, some
measurements of the amplification of an externally applied magnetic field
were carried out \cite{PRL1,MAHYD2,SURVEYS}. Amplification factors up to 25 were 
identified, with
a distinct resonance behaviour at such propeller rotation rates where the 
frequency of the applied field corresponds to the
eigenfrequency  of the dynamo. For almost all rotation rates the measured signal
had only a single component with the frequency of the applied field.
However, at the highest achieved propeller rotation 
rate of 2150 rpm, an additional exponentially increasing mode was 
identified  with a frequency 
different from that of the externally applied field (see Fig. 2).

\begin{figure}[tbp]
\centering
\includegraphics[width=15cm]{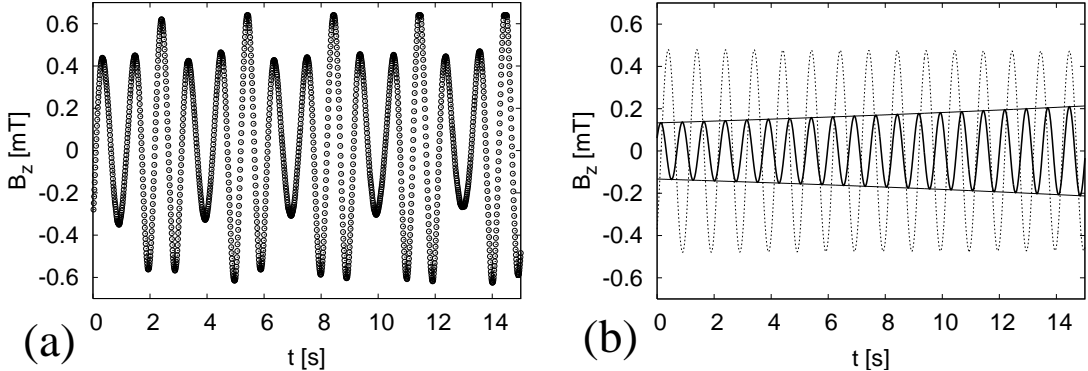} %
\caption{(a) Axial magnetic field measured by a fluxgate sensor (situated close to the 
innermost wall)
at the highest propeller rotation of 2150 rpm on 11 November 1999. 
(b)  Decomposition of this signal into the amplified externally applied
field (dotted line) and the exponentially increasing 
self-excited dynamo eigenmode (full line). }
\end{figure}

Due to some arising technical problems with the seal, this highest 
rotation rate had soon to be reduced
to 1980 rpm, at which the external field was switched off. This allowed the 
observation of a
pure eigenstate, which was already slowly decaying at this lowered  rotation  rate
\cite{PRL1,MAHYD1,SURVEYS}.
In hindsight it can be stated that the eigenfrequencies and growth rates
measured at  these two events
fit perfectly into the many other data that were taken in
later experimental campaigns. 
Hence 11 November 1999 marks the first
observation of an exponentially growing eigenmode in a liquid metal
dynamo experiment.

What was, however,  not achieved in this first experimental campaign was the 
saturated regime
of the dynamo in which the Lorentz forces resulting from  the self-excited
magnetic field reduce the amplitude and modify the structure of the flow velocity
in such a way that the growth rate drops down to zero.
To observe this saturated state it took another 8 months in which
a new seal had to be installed. Then, in July 2000, four runs were 
carried out, partly with, partly without externally
applied magnetic fields, which provided a first  stock of 
growth rate, frequency, 
and spatial structure data of the magnetic eigenfield \cite{PRL2}.

A further step towards the detailed determination of the magnetic eigenfield was done in
June 2002, when several lances with radially spaced Hall sensors were
inserted into the dynamo module.

A somehow frustrating story was the attempt to measure flow 
velocities in a direct manner.
This was, for the first time, tried in February 2003 and reiterated in the
July 2003 and May 2004 campaigns, when three ultrasonic transducers
were installed at the outer wall in order to measure the velocity in the
outermost cylinder of the dynamo. Note that there is no mechanical forcing
of the sodium  in this outer cylinder, and any significant flow there can only result  
from the Lorentz forces due to the magnetic eigenfield. This flow structure was numerically
estimated to be a main rotation with a velocity of the order of 1 m/s, and a
poloidal flow in the form of a double vortex. This flow structure was
experimentally confirmed during some runs in the
May 2004 campaign, while we failed
to study it in more details in further runs. The reason for this failure is 
not completely clear: it seems that, after some short initial phase in which 
data can be collected, the Lorentz force induced flow stirs up a lot of oxides 
in the outer cylinder which then accumulate at the interfaces of the ultrasonic            
transducers 
that may act as cold traps.

Another novelty of the May 2004 campaign was the measurements of pressure data
in the inner dynamo channel by a piezoelectric sensor that was flash 
mounted at the innermost wall. These measurements provided not only
the expected $f^{-7/3}$ behaviour of the pressure fluctuations \cite{AIP}, but also
the velocity modes at the double and the quadruple of the eigenfrequency
of the dynamo which result from the non-axisymmetric parts of the Lorentz forces.

\begin{figure}[tbp]
\centering
\includegraphics[width=15cm]{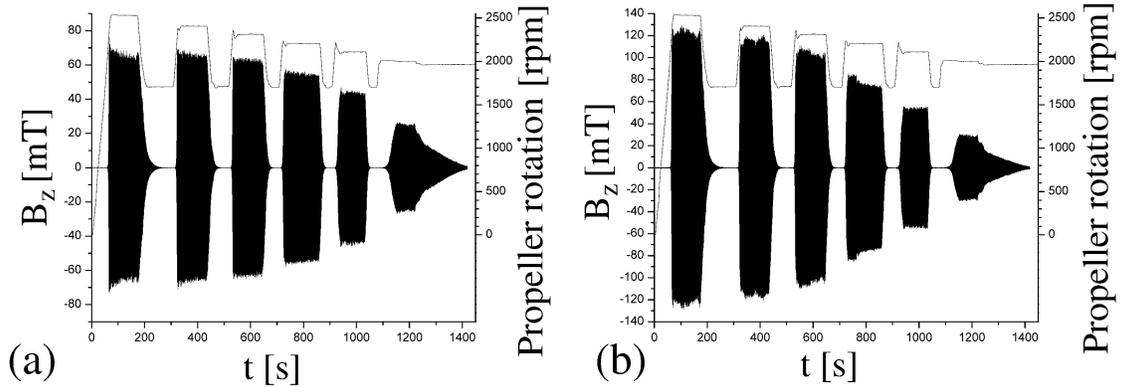} %
\caption{Axial magnetic field and propeller rotation rate during the run 6
of the February/March 2005 campaign, measured by induction coils within the lower
(a) and upper (b) port, close to the innermost wall where the axial magnetic field 
has its maximum.}
\end{figure}

Another attempt to measure velocities in a direct way was undertaken 
in February/March 2005 when a stainless-steel permanent magnet probe for the 
determination of 
flow induced electric potentials  was installed.
In principle, this probe has provided data on the velocity. However, it was
not trivial to disentangle the axial and the azimuthal velocity 
components from the obtained
signals. What was also new  in the February/March 2005 campaign was the installation 
of two
traversing rails with Hall sensors moving in axial direction outside the 
dynamo and induction coils moving in radial direction
within the dynamo module. These traversing sensor rails allowed for the detailed 
determination 
of the spatial structure of the magnetic eigenfield, the results of which will be 
published elsewhere.

Figure 3 shows the axial magnetic field measured by induction coils
inside the upper and the lower port close to the innermost wall
during  the last run
on 1 March 2005. This figure might serve as an example on how 
the magnetic field can be switched on and off at 
will, and on how its amplitude depends on the propeller 
rotation rate.
Comparing Figs. 3a and 3b, a peculiarity of this
dependence becomes visible. Whereas at the upper sensor (Fig. 3b)
the field amplitude increases from 
27 mT for 2000 rpm to 120 mT for 2500 rpm, at the
lower sensor (Fig. 3a) the corresponding
increase is only from  24 mT to 65 mT.
This is a clear indication for a drastic change 
of the field dependence in axial direction with increasing supercriticality of the dynamo, 
which in turn mirrors a significant change of the
axial dependence of the flow.

The most recent campaign was undertaken in July 2007, after that the outworn gliding 
ring seal for the propeller shaft was replaced by a magnetic coupler which had been
designed and produced by SERAS-CNRS in Grenoble. After a few successful runs had been  done 
with this magnetic coupler, the experiment had to be stopped 
due to fact that the belts between the motors and the
propeller shaft were broken.

\section{Summary of main results}

The most significant number for the qualification of a dynamo is the growth rate 
of the magnetic field eigenmode. In the subcritical regime, this negative number 
can either be obtained after switching off an externally applied magnetic field                 
or by lowering the propeller rotation rate from supercritical to subcritical and
tracing again the following exponential decay of the eigenfield. When the dynamo 
is driven into the supercritical state, there are typically a few 
seconds during which the field is still weak enough so that the dynamo can be considered 
as a kinematic one.
Since the Riga dynamo has a complex eigenvalue, there is also a rotation frequency 
of the eigenfield
which can easily be determined both in the kinematic and in the saturated regime.

During the eight experimental campaigns, plenty of growth rate and frequency 
measurements were carried out. When scaled appropriately by the
temperature dependent conductivity of sodium, both quantities turned out to be 
reproducible over the years.  Only a slight change appeared starting with the June 2002
campaign when a lance with sensors was inserted into the central helical flow 
leading to a slight additional braking of the flow and a corresponding slight          
decrease of the growth rate.                                                           

\begin{figure}[tbp]
\centering
\includegraphics[width=15cm]{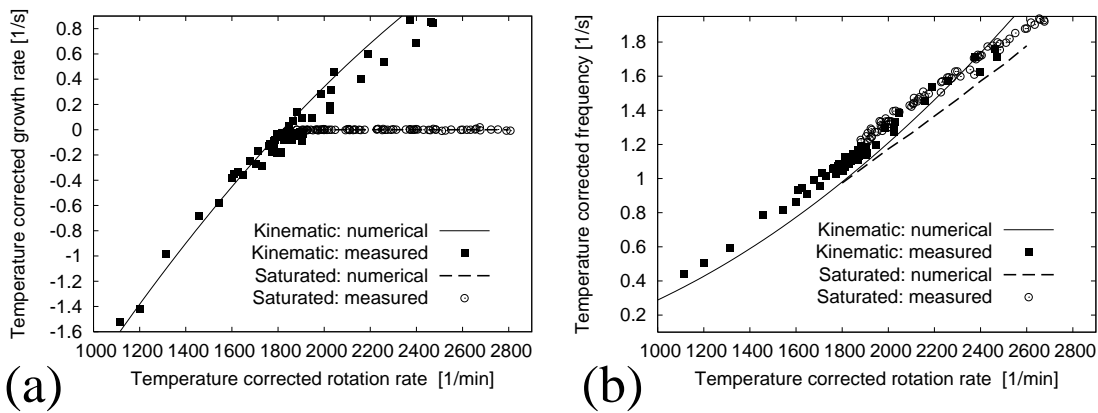} %
\caption{Measured and computed growth rates (a) and frequencies (b) in the
kinematic and saturated regime.  }
\end{figure}

In  Fig. 4, the temperature corrected measurements for the growth rate (Fig 4a) and      
the frequency (Fig. 4b) are shown in comparison with the corresponding
numerical results. The numerical curves in the kinematic regime were obtained with the 
2D solver \cite{KLUWER,PLASMA} and were corrected for the slight effect of the lower      
conductivity of the stainless steel walls
that was estimated  separately  by a 1D solver.
As for the saturation regime we have tried to identify the most important
back-reaction effect within a simple one-dimensional model \cite{MAHYD2,PLASMA}.
While the axial velocity component has to be rather constant from top to bottom
due to mass conservation, the azimuthal component $v_{\phi}$ can be easily 
braked by the
Lorentz forces without any significant pressure increase.
In the inviscid 
approximation, and considering only the $m=0$ mode of the 
Lorentz force, we end up with the  ordinary differential
equation for the Lorentz force induced perturbation  $\delta v_{\phi}(r,z)$
of the azimuthal velocity component:
\begin{eqnarray}
\bar{v}_z(r,z) \frac{\partial}{\partial z} \delta v_{\phi}(r,z) =       
\frac{1}{\mu_0 \rho} [(\nabla \times {\bf{B}}) \times
{\bf{B}}]_{\phi}(r,z)  \;\; .
\label{eq2}
\end{eqnarray} 
In contrast to the procedure described in \cite{MAHYD2,PLASMA} where we had
utilized the measured Joule power at a given
supercriticality
to calibrate the Lorentz force on the r.h.s of Eq. 2, here we solve
Eq. 2 simultaneously with the induction equation. 
Note that Eq. 2 is solved both in the innermost channel where it describes
the downward braking of $v_{\phi}$, as well as in the back-flow channel where it
describes the upward acceleration of $v_{\phi}$. Both effects lead to a reduction of
the differential rotation and hence to a deterioration of the dynamo
capability of the flow.
The validity of this 
self-consistent back-reaction model, which gives automatically a zero 
growth rate, can be judged from the dependence of the
resulting eigenfrequency in Fig. 4b. Actually, we see a quite reasonable correspondence
with the measured data, in particular with respect to the slope of the curve.
A slight jump of the measured eigenfrequencies between the kinematic and the saturated regime
could be attributed to the arising fluid rotation in the outermost cylinder, which
is not described by the back-reaction model according to Eq. 2.
Other criteria for the quality of the back-reaction models are the
correspondence of the resulting amplitude and shape of the 
eigenfield which also turned out to be quite satisfactory.

\begin{figure}[tbp]
\centering
\includegraphics[width=15cm]{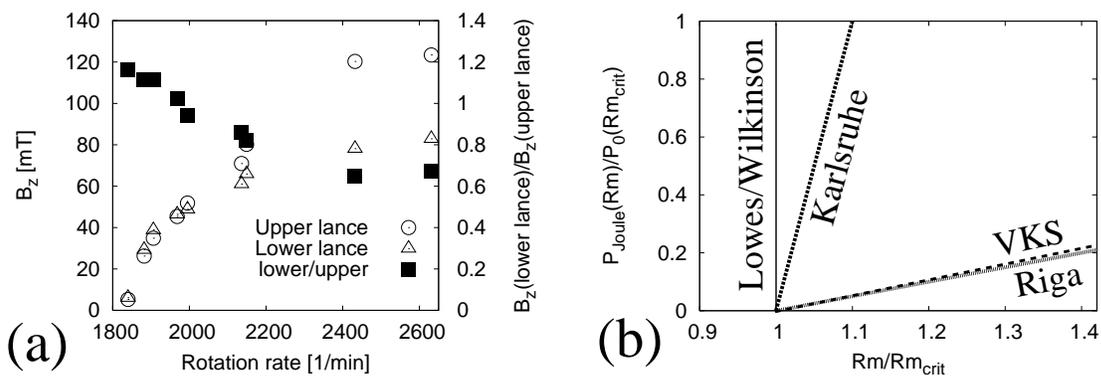}
\caption{(a) Decreasing  ratio of
fields at magnetic field sensors at the lower  and upper lances. (b)
The difference between experimental 
dynamos manifests itself
in the steepness of the Joule power curve beyond the critical
$Rm_c$.
The more flexible  a dynamo is, the easier it can refrain from
producing more magnetic field}
\end{figure}

The braking of the azimuthal velocity component, which accumulates 
downstream from the propeller, 
results in 
a deteriorated self-excitation capability of the flow in the lower parts
of the dynamo and therefore in an upward shift of the magnetic field structure.
This is visible in Fig. 5a which shows 
the measured axial magnetic 
field amplitudes (from the June 2002 campaign)  at the upper
and lower ports and their ratio in 
dependence on the supercriticality. Corresponding dependencies for the              
outer Hall sensors were already documented in Fig. 12 of \cite{SURVEYS}.                 

This deformation of the flow field under the influence of the
Lorentz force of the self-excited magnetic field 
has a strong impact on the 
dependence of dissipated Joule power on the supercriticality. 
In Fig. 5b we try  to compare the 
corresponding curves for the Lowes and Wilkinson experiment \cite{LOWI3}, 
the Karlsruhe
experiment \cite{KARLSRUHE}, the VKS experiment \cite{MONCHAUX}, and the Riga experiment. 
Deliberately, we have given only a schematic plot of these
dependencies since all 
four  curves are not
very accurately known. For the Lowes and Wilkinson case we refer to their
paper \cite{LOWI3},
for the Karlsruhe results we rely on the pressure increase shown 
in the inset of Fig. 4 in \cite{KARLSRUHE}, for VKS we 
took the mean value between the estimates 15 and 20 per cent for 
a supercriticality of
30 per cent \cite{MONCHAUX}, for Riga we 
took the fit $P_{Joule}=(\Omega-1840 \; \mbox{rpm})/(1840 \; \mbox{rpm}) \times 48.4 \; \mbox{k
W}$ of the motor power measurements published in \cite{PLASMA}.
In stark contrast to the 
sharp rise for the Lowes and Wilkinson experiment, but also 
strongly differing from the steep increase in the Karlsruhe 
experiment, the Joule power dependence on supercriticality
in both the VKS and the Riga experiment is very flat. 
For the latter we had already seen that, quite 
different to the back-reaction of a rigid body, the 
sodium flow deforms under the influence of the Lorentz 
forces, and the resulting deterioration of the dynamo
condition makes the growth rate drop down to zero.

\section{Conclusions}

The Riga dynamo experiment, which relies on the concept of the Ponomarenko dynamo,
was the first successful liquid metal experiment in which the critical 
magnetic Reynolds number had been exceeded. Its kinematic regime has been
simulated by a 2D-dimensional finite-difference solver 
with an accuracy of a few percent, and its saturated regime is qualitatively 
well understood by  means of a simplified one-dimensional back-reaction model 
that accounts for the downward braking of the azimuthal component of the
flow. The detailed measurements of the spatial structure of the magnetic eigenfield
in dependence on the supercriticality make the Riga dynamo an ideal test-field
for validating various MHD-turbulence models in contemporary 
three-dimensional RANS model \cite{SASA1,SASA2,SASA3}.

Contrary to what is sometimes written in the literature, the Riga dynamo has
a highly unconstrained flow: the volume fraction in which the fluid flow is directly 
imposed  by the propeller or by guiding blades  is only  10 percent of the
free flow volume behind the propeller. Interestingly, this is the same 10 per cent 
ratio as in the VKS experiment, hence it is not surprising that the slope of the
Joule power curve is approximately the same in both experiments (see Fig. 5b).
In contrast to the VKS experiment, the turbulence                     
level in the Riga experiment is rather low (approximately 8 per cent, depending 
on the position).             
The good correspondence of numerical predictions  
(based on the mean flow) with experimental
data indicates that fluctuations of this level                                 
do not play a significant role in the dynamo mechanism.      

A  drawback of the Riga dynamo could be seen in the fact that it is 
definitely not suited for any
investigation of field reversals since it is  an oscillatory  dynamo
from the very outset. By virtue of its flow topology (a s1t1 in the 
terminology of Dudley and James \cite{DUDLEYJAMES}) there is no chance
to observe any reversal process which, we believe, is 
connected with a transition between
steady and oscillatory eigenvalues of the non-selfadjoint dynamo operator
\cite{STEFANIPRL}.
Such reversal studies have to be left to dynamo experiments with other flow               
topologies, e.g. of the s2t2 type as in the VKS experiment, in which reversals           
were actually observed \cite{BERHANU}.                                                   

The Riga dynamo is still operating. Further improvements of the velocity and 
magnetic field measuring techniques are planned and will possibly help to                 
constrain  turbulence models of flows under the influence of a self-excited 
magnetic field.





\section*{Acknowledgments}
A significant number of people were involved in the construction of the facility
and in carrying out the experimental campaigns. In particular, we are grateful to
Arnis Cifersons, Sergej Dement'ev, Janis Zandarts, Anatolii Zik, 
Michael Christen, Heiko H\"anel, and Gotthard Will.

We thank the Latvian 
Science Council for support under Grants Nos. 96.0276 and 01.0502, the 
Deutsche Forschungsgemeinschaft for support under INK 18/A1-1 and
SFB 609, and the European Commision for support under 
HPRI-CT-2001-50027 and 028679.

\end{document}